\documentclass[10pt,twocolumn,headsepline,DIV=20,paper=A4]{scrartcl}

\pdfoutput=1

\usepackage[UKenglish]{babel}
\usepackage[UKenglish,cleanlook]{isodate}
\usepackage{amsmath}
\usepackage{amssymb}
\usepackage{amsthm}
\usepackage[retainorgcmds]{IEEEtrantools}
\usepackage[bb=boondox]{mathalfa}
\usepackage{titling}
\usepackage{authblk}
\usepackage{abstract}
\usepackage{balance}
\usepackage[square,comma,numbers,sort&compress]{natbib}
\usepackage[colorlinks,linkcolor=blue,citecolor=blue,urlcolor=blue]{hyperref}
\usepackage{quant_defs}

\def\algor{Algorithmica}%
\def\aipconf{AIP Conf.\ Proc.}%
\def\jacm{J.\ ACM}%
\def\natcomm{Nat.\ Commun.}%
\def\natphys{Nature Phys.}%
\def\njp{New J.\ Phys.}%
\def\physics{Physics}%
\def\pra{Phys.\ Rev.\ A}%
\def\prl{Phys.\ Rev.\ Lett.}%
\def\prsa{Proc.\ R.\ Soc.\ A}%
\def\quantic{Quantum Inf.\ Comput.}%
\def\rmp{Rev.\ Mod.\ Phys.}%

\newcommand{\bbar}[1]{\bar{\bar{#1}}}

\newcommand{\rA}{\mathrm{A}}
\newcommand{\rB}{\mathrm{B}}

\newcommand{\rE}{\mathrm{E}}
\newcommand{\rAB}{\mathrm{AB}}

\newcommand{\rBE}{\mathrm{BE}}

\newcommand{\rEE}{\mathrm{EE}}

\newcommand{\rw}{\mathrm{w}}
\newcommand{\rx}{\mathrm{x}}
\newcommand{\ry}{\mathrm{y}}
\newcommand{\rz}{\mathrm{z}}
\newcommand{\rwx}{\mathrm{wx}}
\newcommand{\rxx}{\mathrm{xx}}
\newcommand{\rzx}{\mathrm{zx}}
\newcommand{\rzz}{\mathrm{zz}}

\newcommand{\hilb}{\mathcal{H}}

\newcommand{\0}[0]{^{\vphantom{\prime}}}
\newcommand{\1}[0]{^{\prime}}

\pretitle{\centering\large\bfseries}
\posttitle{\par\vskip 0.5cm}
\preauthor{\centering\small}
\postauthor{\par}
\predate{\centering\small(Dated: }
\postdate{)\par}

\AtBeginDocument{}

\setkomafont{disposition}{\rmfamily\bfseries\boldmath\small}
\setkomafont{section}{\centering\MakeUppercase}
\setkomafont{subsection}{\centering}
\setkomafont{paragraph}{\normalsize\itshape\textmd}

\newcommand{\email}[1]{\thanks{\href{mailto:#1}{#1}}}

\title{Semi device independence of the BB84 protocol}
\author{Erik Woodhead\email{Erik.Woodhead@icfo.es}}
\date{14 February 2017}
\affil{ICFO -- Institut de Ci\`e{}ncies Fot\`o{}niques, The Barcelona
  Institute of Science and Technology, \\
  08860 Castelldefels (Barcelona), Spain}

\newtheorem{lemma}{Lemma}

\begin{document}

\twocolumn[
  \maketitle
  \begin{onecolabstract}
    The BB84 quantum key distribution protocol is semi device independent in
    the sense that it can be shown to be secure if just one of the users'
    devices is restricted to a qubit Hilbert space. Here, we derive an
    analytic lower bound on the asymptotic secret key rate for the
    entanglement-based version of BB84 assuming only that one of the users
    performs unknown qubit POVMs. The result holds against the class of
    collective attacks and reduces to the well known Shor-Preskill key rate
    for correlations corresponding to the ideal BB84 correlations mixed with
    any amount of random noise.
    \vspace{0.5cm}
  \end{onecolabstract}
]
\saythanks

\section{BB84 and device independence}

Quantum key distribution (QKD) \cite{ref:gr2002,ref:sbp2009} protocols allow
cooperating users to generate cryptographic keys in such a way that
unauthorised eavesdropping can be detected. This is achieved by exploiting
features of quantum physics, such as the general inability to measure a
quantum state without disturbing it, in a way that guarantees that any
attempt at eavesdropping on the protocol will introduce detectable errors.

One of a QKD protocol's differentiating features is the degree to which it is
\emph{device independent} \cite{ref:my2004,ref:bhk2005,ref:ab2007}, i.e., the
extent to which the protocol can be proved secure independently of
assumptions about the internal functioning of the devices in the physical
setup. This is of practical interest as device-independent protocols are
intrinsically more robust, ensuring that both unintended and maliciously
introduced implementation faults are detected automatically. Protocols can
range from fully characterised (the exact quantum state preparations and/or
measurements must be known) to fully device independent (security is
established based only on the detection of Bell-nonlocal
\cite{ref:b1964,ref:bc2014} correlations, independently of the mechanism that
produced them). Between these extremes, partially device-independent
protocols have also been proposed in which only some of the devices are fully
characterised \cite{ref:bc2012,ref:bp2012,ref:lcq2012} and in which only a
Hilbert space dimension bound is assumed for the source of quantum states
\cite{ref:pb2011,ref:wp2015}.

The BB84 protocol \cite{ref:bb1984} was originally introduced as a fully
characterised protocol. A commonly considered \emph{prepare-and-measure}
version runs as follows. One user (``Alice'') generates a string of random
bits that she wishes to transmit to another distant user (``Bob''). Alice
sequentially encodes each bit onto one of two corresponding orthogonal $\sz$
eigenstates $\ket{0}$ and $\ket{1}$ which she transmits to Bob. In order to
be able to detect eavesdropping, Alice inserts instances of the $\sx$
eigenstates $\ket{+}$ and $\ket{-}$, with $\ket{\pm} = (\ket{0} \pm \ket{1})
/ \sqrt{2}$, at some random locations in the sequence of quantum states to be
transmitted to Bob. Bob measures most of the states he receives from Alice in
the $\sz = \proj{0} - \proj{1}$ basis and the remaining minority of cases in
the $\sx = \proj{+} - \proj{-}$ basis. Afterwards, the record of cases where
Alice and Bob used mismatched bases (Alice prepared a $\sz$ state and Bob
measured $\sx$ or vice versa) are discarded. The cases where Alice and Bob
both used the $\sx$ basis and a randomly chosen subset of cases where they
both used the $\sz$ basis are used to estimate the $\rx$- and $\rz$-basis
error rates $\delta_{\rx}$ and $\delta_{\rz}$ and then likewise
discarded. Finally, if the error rates are not too high, classical
postprocessing allows a (generally shorter) secret key to be generated with
the relative errors between Alice's and Bob's versions corrected and with any
knowledge of the key by an adversary effectively erased.

There is also an \emph{entanglement-based} version of BB84, in which a
central source prepares and distributes entangled states which Alice, as well
as Bob, measures in the $\sz$ and $\sx$ bases. In this case, the initial
bitstring is obtained from the measurement results rather than from a
separate randomness generation procedure. Since Alice's $\sz$ or $\sx$
measurement can be thought of as effectively preparing a state for Bob
\cite{ref:bbm1992}, there is some equivalence between the two versions of the
protocol. In particular, in both versions, one-way classical postprocessing
allows a secret key to be extracted at an asymptotic rate given by the
Shor-Preskill key rate \cite{ref:sp2000},
\begin{equation}
  \label{eq:shor_preskill_rate}
  r \geq 1 - h(\delta_{\rx}) - h(\delta_{\rz}) \,,
\end{equation}
where $h(x) = - x \log_{2}(x) - (1 - x) \log_{2}(1 - x)$ is the binary
entropy function, depending on the error rates $\delta_{\rx}$ and
$\delta_{\rz}$.

Since its original proposal, it has become apparent that the BB84 protocol
exhibits a significant degree of device independence. BB84 was first found to
be \emph{one-sided device independent}, i.e., the explicit characterisation
of one of the devices can be dropped. This was already indicated by some
early security results \cite{ref:fg1997,ref:bb2002,ref:m2001} for the
prepare-and-measure version of BB84 which do not explicitly depend on Bob's
measurements, and later analyses \cite{ref:kp2003,ref:bc2010} found that the
Shor-Preskill key-rate bound \eqref{eq:shor_preskill_rate} still holds at the
one-sided-device-independent level if Alice's source prepares the $\sz$ and
$\sx$ eigenstates (in the prepare-and-measure version) or just one of the
users measures in the $\sz$ and $\sx$ bases (in the entanglement-based
version).

Recent analyses have started to exploit results from the mismatched bases
cases, which are usually discarded, in order to improve the security
certification \cite{ref:tc2014,ref:cml2016}, and some authors have further
pointed out that this can reduce the level of characterisation required to
just a dimension bound for one of the devices. In Ref.~\cite{ref:gm2014}, it
was first shown that the Shor-Preskill rate still holds if no correlations
are observed in the mismatched bases cases assuming that Alice performs
unknown projective qubit measurements. A similar result was recovered
numerically in Ref.~\cite{ref:gbs2016} for general qubit POVMs on Alice's
side, assuming that Bob also performs qubit measurements. The
prepare-and-measure version of BB84 was also studied numerically in
\cite{ref:yf2014} at a similar level of device independence, where Alice's
source prepares unknown pure qubit states and Bob performs unknown projective
qubit measurements.

Here, we study the BB84 protocol in this \emph{semi-device-independent}
scenario (borrowing the name from \cite{ref:pb2011}), where we assume only
that Alice's device acts on a two-dimensional Hilbert space. The main result
will be an analytic lower bound on the asymptotic secret key rate for the
entanglement-based version of BB84 where we allow Alice's measurements to be
arbitrary qubit POVMs and Bob's measurements are left uncharacterised. The
result holds against the class of collective attacks \cite{ref:bb2002} (i.e.,
assuming that Alice's and Bob's measurements are always performed on the same
entangled state), which is known to imply unconditional security at least if
the measurements are memoryless and if the Hilbert-space dimension is bounded
\cite{ref:ckr2009}.

The qubit device assumption is taken here to mean that Alice's result depends
\emph{only} on the measurement of a qubit state. In particular, similar to
\cite{ref:bqb2014,ref:lb2015}, we assume that Alice's measurement result does
not depend on additional classical information that could also be available
to Bob's device (so-called ``shared randomness'' \cite{ref:pb2011}). This is
necessary as the ideal (entanglement-based) BB84 correlations can be
simulated with two shared classical random bits---a special case of what an
adversary could prepare with a shared classical bit and an entangled qubit
which is completely insecure from a cryptographic perspective. A consequence
is that, unusually for a QKD security result, any (nontrivial) lower bound on
the key rate cannot be a convex function of the probabilities $P(ab \mid uv)$
at this level of device independence.

\section{Scenario and main result}
\label{sec:scenario}

In the entanglement-based version of the BB84 protocol, Alice and Bob share a
state $\rho_{\rAB}$ on some Hilbert space $\hilb_{\rA} \otimes \hilb_{\rB}$,
on which they can perform POVMs $\{M^{(u)}_{0}, M^{(u)}_{1}\}$ and
$\{N^{(v)}_{0}, N^{(v)}_{1}\}$ indexed by measurement choices
$u, v \in \{\rz, \rx\}$ and yielding results $a, b \in \{0, 1\}$ with
probability
\begin{equation}
  \label{eq:quantum_prob}
  P(ab \mid uv) = \Tr \bigsq{(M^{(u)}_{a} \otimes N^{(v)}_{b}) \rho_{\rAB}} \,.
\end{equation}
In the semi-device-independent level of security that we consider, we assume
that $\dim \hilb_{\rA} = 2$. The state $\rho_{\rAB}$ and measurements are
otherwise treated as unknown. Setting $\hat{A}_{u} = M^{(u)}_{0} -
M^{(u)}_{1}$ and $\hat{B}_{v} = N^{(v)}_{0} - N^{(v)}_{1}$, a convenient
summary of the probabilities $P(ab \mid uv)$ that we will use is given by the
eight parameters
\begin{IEEEeqnarray}{rCl}
  A_{u} &=& \avg{\hat{A}_{u} \otimes \id_{\rB}} \,, \\
  B_{v} &=& \avg{\id_{\rA} \otimes \hat{B}_{v}} \,, \\
  E_{uv} &=& \avg{\hat{A}_{u} \otimes \hat{B}_{v}} \,,
\end{IEEEeqnarray}
with $\avg{\;\cdot\;} = \Tr[\;\cdot\; \rho_{\rAB}]$. Note that $E_{\rzz}$ and
$E_{\rxx}$ here are related to the more conventional $\rz$- and $\rx$-basis
error rates $\delta_{\rz}$ and $\delta_{\rx}$ by $E_{uu} = 1 - 2 \delta_{u}$.

The full security analysis of the protocol will be undertaken in the next
section, but it is worth already sketching a result for the special case
where Alice performs rank-one projective measurements since one can be
derived directly from the Shor-Preskill rate. In this scenario, where Alice's
$\rz$ and $\rx$ measurements simply project into orthogonal bases
$\{\ket{0_{\rz}}, \ket{1_{\rz}}\}$ and $\{\ket{0_{\rx}}, \ket{1_{\rx}}\}$,
essentially the only relevant parameter differentiating the measurements is
the Bloch-sphere angle between them. For some suitable basis
$\{\ket{0_{\rw}}, \ket{1_{\rw}}\}$ conjugate to
$\{\ket{0_{\rz}}, \ket{1_{\rz}}\}$, we may write
\begin{equation}
  \hat{A}_{\rx} = \cos(\varphi) \hat{A}_{\rz}
  + \sin(\varphi) \hat{A}_{\rw} \,,
\end{equation}
where $\hat{A}_{\rw} = \proj{0_{\rw}} - \proj{1_{\rw}}$ and $\varphi$ is the
(unknown) Bloch-sphere angle between $\hat{A}_{\rz}$ and
$\hat{A}_{\rx}$. Setting
$E_{\rwx} = \avg{\hat{A}_{\rw} \otimes \hat{B}_{\rx}}$, linearity of the
quantum expectation value implies the relation
\begin{equation}
  \label{eq:Exx_Ezx_Ewx_relation}
  E_{\rxx} = \cos(\varphi) E_{\rzx} + \sin(\varphi) E_{\rwx} \,.
\end{equation}

The conjugate ``$\rw$ basis'' introduced here is useful because the
(one-sided-device-independent) Shor-Preskill key rate applies to
it. Introducing, for convenience, the function
\begin{equation}
  \phi(x) = 1 - \tfrac{1}{2} (1 + x) \log_{2}(1 + x)
  - \tfrac{1}{2} (1 - x) \log_{2}(1 - x)
\end{equation}
(related to the binary entropy by
$\phi(x) = h(\tfrac{1}{2} \pm \tfrac{1}{2} x)$), the Shor-Preskill rate can
be expressed as
\begin{equation}
  \label{eq:shor_preskill_phizw}
  r \geq 1 - \phi(E_{\rwx}) - \phi(E_{\rzz}) \,.
\end{equation}
From here, it is a simple matter to obtain a key-rate bound depending only on
the observed correlations. From the relation \eqref{eq:Exx_Ezx_Ewx_relation}
between the correlators, we obtain
\begin{IEEEeqnarray}{rCl}
  \abs{E_{\rxx}} &\leq& \abs{\cos(\varphi)} \abs{E_{\rzx}}
  + \abs{\sin(\varphi)} \abs{E_{\rwx}} \IEEEnonumber \\
  &\leq& \sqrt{E\du{\rzx}{2} + E\du{\rwx}{2}} \,,
\end{IEEEeqnarray}
which rearranges to
\begin{equation}
  E\du{\rwx}{2} \geq E\du{\rxx}{2} - E\du{\rzx}{2} \,.
\end{equation}
As long as $\abs{E_{\rxx}} \geq \abs{E_{\rzx}}$, this implies the lower bound
\begin{equation}
  \label{eq:bb84_semidev_proj_rate}
  r \geq 1 - \phi \bigro{\sqrt{E\du{\rxx}{2} - E\du{\rzx}{2}}}
  - \phi(E_{\rzz})
\end{equation}
for the key rate.

More generally, it is clear that the key-rate bound
\eqref{eq:bb84_semidev_proj_rate} cannot hold against arbitrary POVMs on
Alice's side. A simple counterexample is that if we allow Alice to perform
the degenerate projective measurement
$\{M^{(\rz)}_{0}, M^{(\rz)}_{1}\} = \{\id_{\rA}, \nil_{\rA}\}$, it is
possible for Alice and Bob to obtain the result $a = b = 0$ deterministically
(which is completely insecure) while observing the correlations
$E_{\rxx} = E_{\rzz} = 1$ and $E_{\rzx} = 0$ (for which
\eqref{eq:bb84_semidev_proj_rate} would imply $r = 1$). Of course, this
particular pathological case is easily detected since Alice and Bob could
notice that they keep getting the same measurement results. In terms of the
parameterisation given above, we thus do not expect
\eqref{eq:bb84_semidev_proj_rate} to still apply if $A_{\rz} = 1$.

There is a significant parameter range in which the rate
\eqref{eq:bb84_semidev_proj_rate} still holds, though. The main result of
this article is that the asymptotic rate \eqref{eq:bb84_semidev_proj_rate}
still applies, at least against collective attacks, if the correlations
satisfy $\abs{E_{\rxx}} > \abs{B_{\rx}}$ and
\begin{equation}
  \label{eq:simple_cond}
  E\du{\rxx}{2} + E\du{\rzx}{2}
  \leq 1 - 2 \abs{A_{\rz} - E_{\rzx} B_{\rx}} + A\du{\rz}{2} \,.
\end{equation}
This is proved in the next section. As a special case, we recover the
Shor-Preskill rate
\begin{equation}
  r \geq 1 - \phi(E_{\rxx}) - \phi(E_{\rzz})
\end{equation}
if there are no correlations in the mismatched bases cases (so that
$E_{\rzx} = 0$) and if
$\abs{B_{\rx}} < \abs{E_{\rxx}} \leq 1 - \abs{A_{\rz}}$; the latter
constraint reduces to $\abs{E_{\rxx}} > 0$ (which is necessary to certify a
nonzero key rate anyway) if Alice's and Bob's marginal results are
equiprobable (so that $A_{\rz} = B_{\rx} = 0$).

In principle, the derivation given in the next section could be pursued
further in order to derive a lower bound for the key rate in the case that
the condition \eqref{eq:simple_cond} is not satisfied. There is an easier way
of getting a result for this case, though. Since the condition
\eqref{eq:simple_cond} and key rate \eqref{eq:bb84_semidev_proj_rate} are
device independent on Bob's side, we can simply apply the result they
\emph{would} imply if Bob's measurement operator $\hat{B}_{\rx}$ were scaled
down to $\lambda \hat{B}_{\rx}$ for some scaling factor $0 \leq \lambda \leq
1$. This way, we can use the modified bound
\begin{equation}
  r \geq 1 - \phi \bigro{\lambda \sqrt{E\du{\rxx}{2} - E\du{\rzx}{2}}}
  - \phi(E_{\rzz}) \,,
\end{equation}
taking for $\lambda$ the highest number between zero and one satisfying
\begin{equation}
  \label{eq:lambda_factor_def}
  \lambda^{2} \bigro{E\du{\rxx}{2} + E\du{\rzx}{2}}
  = 1 - 2 \abs{A_{\rz} - \lambda^{2} E_{\rzx} B_{\rx}} + A\du{\rz}{2} \,.
\end{equation}

\section{Proof of main result}

\subsection*{Problem definition}

In the worst-case scenario, Alice, Bob, and the adversary Eve share a
purification $\ket{\Psi} \in \hilb_{\rA} \otimes \hilb_{\rB} \otimes
\hilb_{\rE}$, prepared by Eve, of the state $\rho_{\rAB}$ responsible for the
observed correlations according to \eqref{eq:quantum_prob}. When Alice
measures $u = \rz$, the system in $\hilb_{\rB} \otimes \hilb_{\rE}$ is
projected to the (unnormalised) state
\begin{equation}
  \rho = \Tr_{\rA} \bigsq{(M^{(\rz)}_{0} \otimes \id_{\rBE}) \Psi}
\end{equation}
or
\begin{equation}
  \rho' = \Tr_{\rA} \bigsq{(M^{(\rz)}_{1} \otimes \id_{\rBE}) \Psi} \,,
\end{equation}
depending, respectively, on whether Alice gets the result $a = 0$ or $a =
1$. (We will in general write, e.g., $\Psi$ as a shorthand for the density
operator $\proj{\Psi}$ associated to some pure state $\ket{\Psi}$.) The
normalisations of these states are related to the probabilities with which
they are prepared according to $\Tr[\rho] = P_{\rA}(0 \mid \rz)$ and
$\Tr[\rho'] = P_{\rA}(1 \mid \rz)$. The correlation between Alice's result
$a$ and the state available to Eve is summarised by the classical-quantum
state
\begin{equation}
  \label{eq:cq_state}
  \tau_{A\rE} = \proj{0} \otimes \rho\0_{\rE}
  + \proj{1} \otimes \rho\1_{\rE} \,,
\end{equation}
in terms of Eve's parts $\rho\0_{\rE} = \Tr_{\rB}[\rho]$ and $\rho\1_{\rE} =
\Tr_{\rB}[\rho']$ of the possible density operators $\rho$ and $\rho'$.

We consider the case where the key is extracted from the $u = v = \rz$
measurement results. In this case, the one-way asymptotic key rate secure
against collective attacks is lower bounded by the Devetak-Winter rate
\cite{ref:dw2005}, which can be expressed as the difference of two entropies
\begin{equation}
  \label{eq:dw_rate}
  r = H(A \mid \rE) - H(A \mid B) \,.
\end{equation}
In \eqref{eq:dw_rate}, $H(A \mid B)$ is the Shannon entropy of Alice's
outcome conditioned on Bob's and can either be computed directly or
approximated by $H(A \mid B) \leq h(\delta_{\rz}) = \phi(E_{\rzz})$.  The
main problem, and the main goal of this section, is to derive a lower bound
for the conditional von Neumann entropy $H(A \mid \rE)$, which is given by
\begin{IEEEeqnarray}{rCl}
  H(A \mid \rE) &=& S(\tau_{A\rE}) - S(\tau_{\rE}) \IEEEnonumber \\
  &=& S(\rho\0_{\rE}) + S(\rho\1_{\rE}) - S(\rho\0_{\rE} + \rho\1_{\rE}) \,,
\end{IEEEeqnarray}
where $S(\rho) = - \Tr[\rho \log_{2}(\rho)]$, when computed on the
classical-quantum state \eqref{eq:cq_state}.

The derivation followed in the remainder of this section uses a few
mathematical tools (two of which are minor restatements of results in
\cite{ref:w2013}) which are presented here as lemmas. Proofs for these are
supplied as appendices to this article.

\subsection*{General proof outline}

The starting point is the following relation for the conditional von Neumann
entropy, which simplifies the problem to that of lower bounding the fidelity
between the marginal states available to Eve.

\begin{lemma}
  \label{lem:HAE_general_bound}
  The conditional von Neumann entropy, computed on the classical-quantum
  state $\proj{0} \otimes \rho\0_{\rE} + \proj{1} \otimes \rho\1_{\rE}$, is
  lower bounded by
  \begin{equation}
    \label{eq:HAE_general_bound}
    H(A \mid \rE) \geq \phi(A_{\rz})
    - \phi \Bigro{\sqrt{A\du{\rz}{2}
        + 4 F(\rho\0_{\rE}, \rho\1_{\rE})^{2}}}
  \end{equation}
  in terms of the fidelity $F(\rho\0_{\rE}, \rho\1_{\rE})$ between
  $\rho\0_{\rE}$ and $\rho\1_{\rE}$. Furthermore, for fixed $F(\rho\0_{\rE},
  \rho\1_{\rE})$, the right-hand side of \eqref{eq:HAE_general_bound} is
  convex in $A_{\rz}$ and is minimised with $A_{\rz} = 0$.
\end{lemma}

Here, we take the fidelity to be defined by
$F(\rho, \sigma) = \trnorm{\sqrt{\rho} \sqrt{\sigma}}$, where
$\trnorm{A} = \Tr[\abs{A}] = \Tr[\sqrt{A^{\dagger} A}]$ denotes the trace
norm of an operator $A$, for (generally unnormalised) density operators
$\rho$ and $\sigma$. Note that the minimisation of
\eqref{eq:HAE_general_bound} at $A_{\rz} = 0$ allows the bound for the von
Neumann entropy to be simplified to
\begin{equation}
  \label{eq:HAE_simple_bound}
  H(A \mid \rE) \geq  1 - \phi \bigro{2 F(\rho\0_{\rE}, \rho\1_{\rE})} \,,
\end{equation}
though this step is optional, since $A_{\rz}$ is an observed parameter.

The approach we follow involves reducing the problem to considering pure
states. To this end, we introduce orthonormal bases
$\{\ket{0_{u}}, \ket{1_{u}}\}$, $u \in \{\rz, \rx\}$, in which Alice's (qubit
Hermitian) POVM elements $M^{(u)}_{a}$ are diagonal. In these bases, Alice's
POVMs can be expressed as convex sums
\begin{IEEEeqnarray}{rCl}
  \label{eq:povm_decomp}
  \{M^{(u)}_{0}, M^{(u)}_{1}\}
  &=& m^{(u)}_{1} \{0_{u}, 1_{u}\} + m^{(u)}_{2} \{1_{u}, 0_{u}\}
  \IEEEnonumber \\
  &&+\> m^{(u)}_{3} \{\id_{\rA}, \nil_{\rA}\}
  + m^{(u)}_{4} \{\nil_{\rA}, \id_{\rA}\} \IEEEeqnarraynumspace
\end{IEEEeqnarray}
of the four projective measurements $\{0_{u}, 1_{u}\}$, $\{1_{u}, 0_{u}\}$,
$\{\id_{\rA}, \nil_{\rA}\}$, and $\{\nil_{\rA}, \id_{\rA}\}$ for convex
coefficients satisfying $m^{(u)}_{i} \geq 0$ and $\sum_{i} m^{(u)}_{i} =
1$. (Here, $0_{u}$ and $1_{u}$ are shorthand for $\proj{0_{u}}$ and
$\proj{1_{u}}$, and $\id_{\rA}$ and $\nil_{\rA}$ denote the identity and null
operators on $\hilb_{\rA}$.)

Concentrating on the $\rz$ measurement, we can express the entangled state as
\begin{equation}
  \label{eq:psi_z_decomp}
  \ket{\Psi} = \ket{0_{\rz}} \ket{\alpha} + \ket{1_{\rz}} \ket{\alpha'}
\end{equation}
for (unnormalised and not necessarily orthogonal) states $\ket{\alpha},
\ket{\alpha'} \in \hilb_{\rB} \otimes \hilb_{\rE}$. The fidelity between
Eve's parts $\alpha\0_{\rE}$ and $\alpha\1_{\rE}$ of the states
$\ket{\alpha}$ and $\ket{\alpha'}$ introduced this way can, according to the
following relation, be bounded in terms of an operator $W_{\rB}$ on Bob's
Hilbert space.

\begin{lemma}
  \label{lem:fidel_trnormW}
  The fidelity between Eve's partial traces $\alpha\0_{\rE}$ and
  $\alpha\1_{\rE}$ of the pure states $\ket{\alpha}$ and $\ket{\alpha'}$
  satisfies
  \begin{equation}
    2 F(\alpha\0_{\rE}, \alpha\1_{\rE}) \geq \trnorm{W_{\rB}} \,,
  \end{equation}
  where $W_{\rB} = \Tr_{\rE}[W]$ and
  $W = \trans{\alpha}{\alpha'} + \trans{\alpha'}{\alpha}$.
\end{lemma}

We approach the problem of lower bounding $\trnorm{W_{\rB}}$ in the following
way. Similar to \eqref{eq:psi_z_decomp}, we express the entangled state as
\begin{equation}
  \label{eq:psi_x_decomp}
  \ket{\Psi} = \ket{0_{\rx}} \ket{\beta} + \ket{1_{\rx}} \ket{\beta'}
\end{equation}
for the $u = \rx$ measurement. In an appropriate phase convention, the
diagonalising bases are related by
\begin{IEEEeqnarray}{rCl}
  \ket{0_{\rz}} &=& \cos(\tfrac{\varphi}{2}) \ket{0_{\rx}}
  - \sin(\tfrac{\varphi}{2}) \ket{1_{\rx}} \,, \\
  \ket{1_{\rz}} &=& \sin(\tfrac{\varphi}{2}) \ket{0_{\rx}}
  + \cos(\tfrac{\varphi}{2}) \ket{1_{\rx}}
\end{IEEEeqnarray}
for some angle $\varphi$. From this and requiring that
\eqref{eq:psi_z_decomp} and \eqref{eq:psi_x_decomp} are the same state, we
extract
\begin{IEEEeqnarray}{rCl}
  \label{eq:beta0_alpha_phi}
  \ket{\beta} &=& \cos(\tfrac{\varphi}{2}) \ket{\alpha}
  + \sin(\tfrac{\varphi}{2}) \ket{\alpha'} \,, \\
  \label{eq:beta1_alpha_phi}
  \ket{\beta'} &=& -\sin(\tfrac{\varphi}{2}) \ket{\alpha}
  + \cos(\tfrac{\varphi}{2}) \ket{\alpha'} \,.
\end{IEEEeqnarray}
Introducing the correlators
\begin{IEEEeqnarray}{rCl}
  \bbar{E}_{\rzx} &=& \Tr \bigsq{
    \hat{B}_{\rx} (\alpha\0_{\rB} - \alpha\1_{\rB})} \,, \\
  \bbar{E}_{\rxx} &=& \Tr \bigsq{
    \hat{B}_{\rx} (\beta\0_{\rB} - \beta\1_{\rB})}
\end{IEEEeqnarray}
for the pure states and
\begin{equation}
  \bbar{E}_{\rwx} = \Tr \bigsq{\hat{B}_{\rx} W_{\rB}}
\end{equation}
for the operator $W$ appearing in Lemma~\ref{lem:fidel_trnormW}, the
relations \eqref{eq:beta0_alpha_phi} and \eqref{eq:beta1_alpha_phi} imply
\begin{equation}
  \bbar{E}_{\rxx} = \cos(\varphi) \bbar{E}_{\rzx}
  + \sin(\varphi) \bbar{E}_{\rwx} \,,
\end{equation}
and applying the Cauchy-Schwarz inequality and rearranging, we obtain
\begin{equation}
  \label{eq:bbarEwxz}
  \bbar{E}\du{\rwx}{2}
  \geq \bbar{E}\du{\rxx}{2} - \bbar{E}\du{\rzx}{2} \,,
\end{equation}
similar to the outline of the previous section. Finally, since
$\hat{B}_{\rx}$ is the difference of two POVM elements, it satisfies the
operator inequalities $-\id_{\rB} \leq \hat{B}_{\rx} \leq \id_{\rB}$; this
allows $\bbar{E}_{\rwx}$ to be used as a lower bound on the trace norm
$\trnorm{W_{\rB}}$ of $W_{\rB}$:
\begin{equation}
  \bbar{E}_{\rwx} = \Tr \bigsq{\hat{B}_{\rx} W_{\rB}}
  \leq \trnorm{W_{\rB}} \norm{\hat{B}_{\rx}}_{\infty}
  \leq \trnorm{W_{\rB}} \,,
\end{equation}
from which we finally obtain
\begin{equation}
  \label{eq:fidel_alpha_bound}
  4 F(\alpha\0_{\rE}, \alpha\1_{\rE})^{2}
  \geq \bbar{E}\du{\rxx}{2} - \bbar{E}\du{\rzx}{2} \,.
\end{equation}

The remaining problem is to convert \eqref{eq:fidel_alpha_bound} into a lower
bound on $F(\rho\0_{\rE}, \rho\1_{\rE})$ depending on the observed parameters
$A_{u}$, $B_{v}$, and $E_{uv}$ which can be used in
Lemma~\ref{lem:HAE_general_bound} (or \eqref{eq:HAE_simple_bound}). Part of
the problem is to relate these parameters to the pure-state versions
$\bbar{E}_{\rxx}$ and $\bbar{E}_{\rzx}$ appearing in
\eqref{eq:fidel_alpha_bound}. From the POVM
decomposition \eqref{eq:povm_decomp} we can deduce
\begin{equation}
  \label{eq:E_uv_decomp}
  E_{uv} = (m^{(u)}_{1} - m^{(u)}_{2}) \bbar{E}_{uv}
  + (m^{(u)}_{3} - m^{(u)}_{4}) B_{v} \,,
\end{equation}
which will allow the $\bbar{E}_{uv}$s to be related to the $E_{uv}$s and
$B_{v}$s. For the $\rz$ measurement, we will also need to be able to relate
the fidelity $F(\alpha\0_{\rE}, \alpha\1_{\rE})$ in
\eqref{eq:fidel_alpha_bound} to $F(\rho\0_{\rE}, \rho\1_{\rE})$. For this, we
will need the following general bound for the fidelity between mixtures of
two states.

\begin{lemma}
  \label{lem:fidel_trcoeffs}
  Let $\rho$, $\sigma$, $\tau_{0}$, and $\tau_{1}$ be (not necessarily
  normalised) density operators related by
  \begin{IEEEeqnarray}{rCl}
    \rho &=& p_{0} \tau_{0} + p_{1} \tau_{1} \,, \\
    \sigma &=& q_{0} \tau_{0} + q_{1} \tau_{1}
  \end{IEEEeqnarray}
  for parameters $p_{0}, p_{1}, q_{0}, q_{1} \geq 0$. Then,
  \begin{IEEEeqnarray}{rCl}
    F(\rho, \sigma)^{2} &\geq& \bigro{\sqrt{p_{0} q_{0}} \trnorm{\tau_{0}}
      + \sqrt{p_{1} q_{1}} \trnorm{\tau_{1}}}^{2} \IEEEnonumber \\
    &&+\> \bigro{\sqrt{p_{0} q_{1}} - \sqrt{p_{1} q_{0}}}^{2}
      F(\tau_{0}, \tau_{1})^{2} \,.
  \end{IEEEeqnarray}
\end{lemma}

\subsection*{Alice's $\rx$ POVM}

The $u = \rx$ measurement is the simplest to handle, since it is not used for
key generation, so we deal with it first. Rewriting the decomposition
\eqref{eq:E_uv_decomp} for $E_{\rxx}$ as
\begin{equation}
  \label{eq:Exx_povm_decomp}
  E_{\rxx} = \lambda \bbar{E}_{\rxx}
  + \mu B_{\rx} \,,
\end{equation}
with $\lambda = m^{(\rx)}_{1} - m^{(\rx)}_{2}$ and
$\mu = m^{(\rx)}_{3} - m^{(\rx)}_{4}$, the triangle inequality and the
constraint $\abs{\mu} \leq 1 - \abs{\lambda}$ together imply
\begin{equation}
  \abs{E_{\rxx}} \leq \abs{\lambda} \abs{\bbar{E}_{\rxx}}
  + (1 - \abs{\lambda}) \abs{B_{\rx}} \,,
\end{equation}
which rearranges to
\begin{equation}
  \label{eq:delta_bbE_E_ineq}
  \abs{\lambda} \bigro{\abs{\bbar{E}_{\rxx}} - \abs{E_{\rxx}}}
  \geq (1 - \abs{\lambda})
  \bigro{\abs{E_{\rxx}} - \abs{B_{\rx}}} \,.
\end{equation}
If $\abs{E_{\rxx}} > \abs{B_{\rx}}$ then the only way that
\eqref{eq:delta_bbE_E_ineq} can be satisfied is if $\abs{\lambda} > 0$ and if
$\abs{\bbar{E}_{\rxx}} \geq \abs{E_{\rxx}}$. In this case $E_{\rxx}$ can
safely be substituted in place of $\bbar{E}_{\rxx}$ in the pure-state
fidelity bound \eqref{eq:fidel_alpha_bound}. Otherwise, it is perfectly
possible for the $\rx$ measurement POVM decomposition
\eqref{eq:Exx_povm_decomp} to be satisfied with $\bbar{E}_{\rxx} = 0$. In the
following, we will assume that $\abs{E_{\rxx}} > \abs{B_{\rx}}$, since
\eqref{eq:fidel_alpha_bound} becomes trivial otherwise.

\subsection*{Alice's $\rz$ POVM}

The POVM decomposition \eqref{eq:povm_decomp} implies that the states $\rho$
and $\rho'$ prepared on $\hilb_{\rB} \otimes \hilb_{\rE}$ are related to
$\alpha$ and $\alpha'$ by
\begin{IEEEeqnarray}{rCl}
  \label{eq:rho0_alpha_decomp}
  \rho &=& m^{(\rz)}_{1} \alpha + m^{(\rz)}_{2} \alpha'
  + m^{(\rz)}_{3} (\alpha + \alpha') \,, \\
  \label{eq:rho1_alpha_decomp}
  \rho' &=& m^{(\rz)}_{1} \alpha' + m^{(\rz)}_{2} \alpha
  + m^{(\rz)}_{4} (\alpha + \alpha') \,.
\end{IEEEeqnarray}
In general, the decomposition \eqref{eq:povm_decomp} for POVMs is not unique,
so we have some freedom to choose a decomposition which will simplify the
problem of turning the fidelity bound
\begin{equation}
  \label{eq:fidel_purestates}
  4 F(\alpha\0_{\rE}, \alpha\1_{\rE})^{2}
  \geq E\du{\rxx}{2} - \bbar{E}\du{\rzx}{2}
\end{equation}
into a lower bound for $F(\rho\0_{\rE}, \rho\1_{\rE})$ depending on observed
parameters $A_{u}$, $B_{v}$, and $E_{uv}$. Specifically, the identity
\begin{equation}
  \{0_{\rz}, 1_{\rz}\} + \{1_{\rz}, 0_{\rz}\}
  = \{\id_{\rA}, \nil_{\rA}\} + \{\nil_{\rA}, \id_{\rA}\}
\end{equation}
implies that one of the POVMs $\{\id_{\rA}, \nil_{\rA}\}$ or $\{\nil_{\rA},
\id_{\rA}\}$ can always be eliminated, meaning we can assume that one of
$m^{(\rz)}_{3}$ and $m^{(\rz)}_{4}$ in \eqref{eq:povm_decomp} is zero without
loss of generality.

We proceed in two steps, first considering mixtures of the measurements
$\{0_{\rz}, 1_{\rz}\}$ and $\{1_{\rz}, 0_{\rz}\}$, before accounting for a
contribution from one of the measurements $\{\id_{\rA}, \nil_{\rA}\}$ or
$\{\nil_{\rA}, \id_{\rA}\}$. In anticipation, and assuming a contribution
from $\{\nil_{\rA}, \id_{\rA}\}$ for example, we reexpress
\eqref{eq:rho0_alpha_decomp} and \eqref{eq:rho1_alpha_decomp} as
\begin{IEEEeqnarray}{rCl}
  \label{eq:rho0_pq_alpha}
  \rho &=& p (q \alpha + q' \alpha') \,, \\
  \label{eq:rho1_pq_alpha}
  \rho' &=& p (q' \alpha + q \alpha') + p' (\alpha + \alpha') \,,
\end{IEEEeqnarray}
where the nonnegative parameters $p$, $p'$, $q$, $q'$ are related to the
$m^{(\rz)}_{i}$s by $p = m^{(\rz)}_{1} + m^{(\rz)}_{2}$,
$p' = m^{(\rz)}_{4}$, $p q = m^{(\rz)}_{1}$, and $p q' = m^{(\rz)}_{2}$, and
satisfy $p + p' = q + q' = 1$.

For the contribution from $\{0_{\rz}, 1_{\rz}\}$ and $\{1_{\rz}, 0_{\rz}\}$,
we set
\begin{IEEEeqnarray}{rCl}
  \bar{\rho} &=& q \alpha + q' \alpha' \,, \\
  \bar{\rho}' &=& q' \alpha + q \alpha' \,,
\end{IEEEeqnarray}
and, applying Lemma~\ref{lem:fidel_trcoeffs} and the pure-state fidelity
bound \eqref{eq:fidel_purestates}, we have
\begin{IEEEeqnarray}{rCl}
  4 F(\bar{\rho}\0_{\rE}, \bar{\rho}\1_{\rE})^{2}
  &\geq& 4 q q' + (q - q')^{2} 4 F(\alpha\0_{\rE}, \alpha\1_{\rE})^{2}
  \IEEEnonumber \\
  &\geq& 4 q q' + (q - q')^{2}
  \bigro{E\du{\rxx}{2} - \bbar{E}\du{\rzx}{2}} \,.
\end{IEEEeqnarray}
Introducing the correlator
\begin{equation}
  \bar{E}_{\rzx} = \Tr \bigsq{
    \hat{B}_{\rx} (\bar{\rho}\0_{\rB} - \bar{\rho}\1_{\rB})} \,,
\end{equation}
related to $\bbar{E}_{\rzx}$ by $\bar{E}_{\rzx} = (q - q') \bbar{E}_{\rzx}$,
and using that $4 q q' \geq 4 q q' E\du{\rxx}{2}$,
\begin{IEEEeqnarray}{rCl}
   4 F(\bar{\rho}\0_{\rE}, \bar{\rho}\1_{\rE})^{2}
   &\geq& \bigro{4 q q' + (q - q')^{2}} E\du{\rxx}{2}
   - \bar{E}\du{\rzx}{2} \IEEEnonumber \\
   &=& (q + q')^{2} E\du{\rxx}{2}
   - \bar{E}\du{\rzx}{2} \,,
\end{IEEEeqnarray}
or
\begin{equation}
  \label{eq:fidel_bitflip}
  4 F(\bar{\rho}\0_{\rE}, \bar{\rho}\1_{\rE})^{2}
  \geq E\du{\rxx}{2} - \bar{E}\du{\rzx}{2} \,,
\end{equation}
which shows that allowing mixtures of the measurements $\{0_{\rz}, 1_{\rz}\}$
and $\{1_{\rz}, 0_{\rz}\}$ alone will not affect the key-rate formula.

Finally, we account for the effect of a contribution from one of the
degenerate measurements $\{\id_{\rA}, \nil_{\rA}\}$ or $\{\nil_{\rA},
\id_{\rA}\}$. Assuming first a contribution from $\{\nil_{\rA}, \id_{\rA}\}$,
according to \eqref{eq:rho0_pq_alpha} and \eqref{eq:rho1_pq_alpha} and using
that $\bar{\rho} + \bar{\rho}' = \alpha + \alpha'$, $\rho$ and $\rho'$ are
related to the states $\bar{\rho}$ and $\bar{\rho}'$ defined above by
\begin{IEEEeqnarray}{rCl}
  \label{eq:r0_b1}
  \rho &=& p \bar{\rho} \,, \\
  \label{eq:r1_b1}
  \rho' &=& p' \bar{\rho} + \bar{\rho}' \,.
\end{IEEEeqnarray}
Applying Lemma~\ref{lem:fidel_trcoeffs} again,
\begin{equation}
  F(\rho\0_{\rE}, \rho\1_{\rE})^{2} \geq p p' \norm{\bar{\rho}}\du{1}{2}
  + p F(\bar{\rho}\0_{\rE}, \bar{\rho}\1_{\rE})^{2} \,.
\end{equation}
Inserting the lower bound \eqref{eq:fidel_bitflip} for $F(\bar{\rho}\0_{\rE},
\bar{\rho}\1_{\rE})$ and recognising that
\begin{equation}
  p \trnorm{\bar{\rho}} = \trnorm{\rho}
  = P_{\rA}(0 \mid \rz) = (1 + A_{\rz}) / 2 \,,
\end{equation}
the lower bound for $F(\rho\0_{\rE}, \rho\1_{\rE})$ becomes
\begin{equation}
  \label{eq:fidel_pAEbar_b1}
  4 F(\rho\0_{\rE}, \rho\1_{\rE})^{2}
  \geq \bigro{\tfrac{1}{p} - 1} (1 + A_{\rz})^{2}
  + p E\du{\rxx}{2} - p \bar{E}\du{\rzx}{2} \,.
\end{equation}
The observed parameters
\begin{equation}
  E_{\rzx} = \Tr \bigsq{\hat{B}_{\rx} (\rho\0_{\rB} - \rho\1_{\rB})}
\end{equation}
and
\begin{equation}
  B_{\rx} = \Tr \bigsq{\hat{B}_{\rx} (\rho\0_{\rB} + \rho\1_{\rB})}
  = \Tr \bigsq{\hat{B}_{\rx} (\bar{\rho}\0_{\rB} + \bar{\rho}\1_{\rB})}
\end{equation}
are related to $\bar{E}_{\rzx}$ by
\begin{equation}
  E_{\rzx} = p \bar{E}_{\rzx} - p' B_{\rx} \,.
\end{equation}
Rearranging for $\bar{E}_{\rzx}$ and inserting in
\eqref{eq:fidel_pAEbar_b1}, we obtain
\begin{IEEEeqnarray}{rCl}
  \label{eq:fidel_pAE_b1}
  4 F(\rho\0_{\rE}, \rho\1_{\rE})^{2}
  &\geq& \bigro{\tfrac{1}{p} - 1} (1 + A_{\rz})^{2}
  + p E\du{\rxx}{2} \IEEEnonumber \\
  &&-\> p \Bigro{\tfrac{1}{p} E_{\rzx}
    + \bigro{\tfrac{1}{p} - 1} B_{\rx}}^{2} \,,
\end{IEEEeqnarray}
or, subtracting $E\du{\rxx}{2} - E\du{\rzx}{2}$ from both sides,
\begin{IEEEeqnarray}{rl}
  \IEEEeqnarraymulticol{2}{l}{\label{eq:fidel_diff_b1}
    4 F(\rho\0_{\rE}, \rho\1_{\rE})^{2} - (E\du{\rxx}{2} - E\du{\rzx}{2})}
  \IEEEnonumber \\
  \qquad \geq \bigro{\tfrac{1}{p} - 1} \Bigsq{
    (1 + A_{\rz})^{2} - p(E\du{\rxx}{2} - B\du{\rx}{2}) &
    \IEEEnonumber \\
    -\> (E_{\rzx} + B_{\rx})^{2} &} \,.
\end{IEEEeqnarray}
By following similar reasoning starting from the decomposition
\begin{IEEEeqnarray}{rCl}
  \label{eq:r0_b0}
  \rho &=& \bar{\rho} + p' \bar{\rho}' \,, \\
  \label{eq:r1_b0}
  \rho' &=& p \bar{\rho}' \,,
\end{IEEEeqnarray}
assuming a contribution from $\{\id_{\rA}, \nil_{\rA}\}$ instead of
$\{\nil_{\rA}, \id_{\rA}\}$, we obtain the same result as
\eqref{eq:fidel_diff_b1} except with the sign changes $A_{\rz} \to -A_{\rz}$
and $B_{\rx} \to - B_{\rx}$. The worst of the two bounds obtained this way is
\begin{IEEEeqnarray}{rl}
  \IEEEeqnarraymulticol{2}{l}{\label{eq:fidel_diff}
    4 F(\rho\0_{\rE}, \rho\1_{\rE})^{2} - (E\du{\rxx}{2} - E\du{\rzx}{2})}
  \IEEEnonumber \\
  \qquad \geq \bigro{\tfrac{1}{p} - 1} \Bigsq{&
    1 - 2 \abs{A_{\rz} - E_{\rzx} B_{\rx}} + A\du{\rz}{2} \IEEEnonumber \\
    &-\> p (E\du{\rxx}{2} - B\du{\rx}{2})
    - (E\du{\rzx}{2} + B\du{\rx}{2})} \,. \IEEEeqnarraynumspace
\end{IEEEeqnarray}
The multiplicative factor $1/p - 1$ is nonnegative, so the right-hand side of
\eqref{eq:fidel_diff} is nonnegative if
\begin{IEEEeqnarray}{c}
  \label{eq:simple_cond_p}
  p (E\du{\rxx}{2} - B\du{\rx}{2}) + (E\du{\rzx}{2} + B\du{\rx}{2})
  \IEEEnonumber \\
  \qquad \leq 1 - 2 \abs{A_{\rz} - E_{\rzx} B_{\rx}} + A\du{\rz}{2} \,.
\end{IEEEeqnarray}
Finally, since we are assuming $\abs{E_{\rxx}} > \abs{B_{\rx}}$, the term
$p (E\du{\rxx}{2} - B\du{\rx}{2})$ is nonnegative and is maximised with $p$ =
1. This implies that \eqref{eq:simple_cond_p} is satisfied for all $p \leq 1$
if it is satisfied for $p = 1$, i.e., if
\begin{equation}
  \label{eq:fidel_simple_cond}
  E\du{\rxx}{2} + E\du{\rzx}{2}
  \leq 1 - 2 \abs{A_{\rz} - E_{\rzx} B_{\rx}} + A\du{\rz}{2} \,,
\end{equation}
which is the condition given in the previous section. If this condition is
met then the lower bound
\begin{equation}
  \label{eq:fidel_Exx_Ezx}
  4 F(\rho\0_{\rE}, \rho\1_{\rE})^{2} \geq E\du{\rxx}{2} - E\du{\rzx}{2}
\end{equation}
can be used for the fidelity in Lemma~\ref{lem:HAE_general_bound}.

\section{Conclusion}

The preceding section proves that the key rate asymptotically secure against
collective attacks for BB84 is lower bounded by
\begin{equation}
  \label{eq:r_higher_rate}
  r \geq \phi(A_{\rz}) - \phi \bigro{\sqrt{A\du{\rz}{2}
      + E\du{\rxx}{2} - E\du{\rzx}{2}}}
  - \phi(E_{\rzz})
\end{equation}
if $\abs{E_{\rxx}} > \abs{B_{\rx}}$ and if the condition
\eqref{eq:fidel_simple_cond} is satisfied. This is never less than the
simpler bound \eqref{eq:bb84_semidev_proj_rate} claimed in
section~\ref{sec:scenario}. If \eqref{eq:fidel_simple_cond} is not satisfied,
device independence on Bob's side still allows the main result to be used
with the replacements $E_{\rxx} \to \lambda E_{\rxx}$ and
$E_{\rzx} \to \lambda E_{\rzx}$, with the scaling factor $\lambda$ determined
by \eqref{eq:lambda_factor_def} above.  Together, these give a general
semi-device-independent security result for the BB84 protocol against
collective (and possibly \cite{ref:ckr2009} more general) attacks. The
traditional set of assumptions used to prove the security of the BB84
protocol can thus be relaxed to a significant degree. It is still necessary
to trust that one of the users' measurements are restricted to a
two-dimensional Hilbert space, but exact knowledge of the measurements beyond
this is not required.

In the scenario considered, aside from the qubit restriction on Alice's side,
Alice's and Bob's measurements were allowed to be arbitrary POVMs. One could
go further, similar to \cite{ref:bqb2014,ref:lb2015}, and imagine that Eve
may have more detailed knowledge of the measurements. Specifically, the
approach followed in this article could probably be modified to allow Eve to
know the indices $i$ and $j$ in decompositions of the form
$M^{(u)}_{a} = \sum_{i} p_{i} M^{(u)}_{a; i}$ and
$N^{(v)}_{b} = \sum_{j} q_{j} N^{(v)}_{b; j}$ for the POVM elements, although
the resulting key rate will probably not include the Shor-Preskill rate as a
special case if the adversary is granted this extra power.

Finally, the main result was derived for the entanglement-based version of
BB84. It is likely that a similar result should hold for the
prepare-and-measure BB84 variant assuming a source which is restricted to
emitting qubit states, which was tested in a recent implementation
\cite{ref:xw2015}. Adapting the approach followed here for the
prepare-and-measure scenario is thus an obvious problem for future work.

\section*{Acknowledgements}

Stefano Pironio suggested it would be interesting to study BB84 as a
semi-device-independent protocol back in early 2013 and offered helpful
criticism of a draft of this article. This work is supported by the Spanish
MINECO (Severo Ochoa grant SEV-2015-0522 and FOQUS FIS2013-46768-P), the
Generalitat de Catalunya (SGR 875), the Fundaci\'o{} Privada Cellex, and the
EU project QITBOX.

\bibliography{bb84_qubit}

\appendix

\subsection*{Proof of Lemma~\ref{lem:HAE_general_bound}}

The conditional von Neumann entropy satisfies
$H(A \mid \rE) \geq H(A \mid \rEE')$ for any extension
$\hilb_{\rE} \otimes \hilb_{\rE'}$ of Eve's Hilbert space $\hilb_{\rE}$. We
use this to replace the (unnormalised) density operators $\rho\0_{\rE}$ and
$\rho\1_{\rE}$ appearing in the classical-quantum state \eqref{eq:cq_state}
with purifications $\ket{\psi}$ and $\ket{\psi'}$; by Uhlmann's theorem
(which still holds for unnormalised states), these can be chosen such that
$\braket{\psi}{\psi'} = F(\rho\0_{\rE}, \rho\1_{\rE})$. We this way obtain
\begin{IEEEeqnarray}{rCl}
  H(A \mid \rE) &\geq& S(\psi) + S(\psi')
  - S(\psi + \psi') \IEEEnonumber \\
  &=& h \bigro{P_{\rA}(0 \mid \rz)} - h(\lambda_{+}) \,,
\end{IEEEeqnarray}
where
\begin{equation}
  \lambda_{\pm} = \tfrac{1}{2} \pm \tfrac{1}{2} \sqrt{(\trnorm{\psi} -
  \trnorm{\psi'})^{2} + 4 F(\rho\0_{\rE}, \rho\1_{\rE})^{2}}
\end{equation}
are the eigenvalues of $\psi + \psi'$. Recognising that
\begin{IEEEeqnarray}{rCl}
  \trnorm{\psi} - \trnorm{\psi'} &=& \trnorm{\rho} - \trnorm{\rho'}
  \IEEEnonumber \\
  &=& P_{\rA}(0 \mid \rz) - P_{\rA}(1 \mid \rz) \IEEEnonumber \\
  &=& A_{\rz} \,,
\end{IEEEeqnarray}
we obtain
\begin{equation}
  \label{eq:HAE_bound_appendix}
  H(A \mid \rE) \geq \phi(A_{\rz})
  - \phi \Bigro{\sqrt{A\du{\rz}{2}
      + 4 F(\rho\0_{\rE}, \rho\1_{\rE})^{2}}} \,,
\end{equation}
which is the lower bound claimed in the statement of
Lemma~\ref{lem:HAE_general_bound}.

The right-hand side of \eqref{eq:HAE_bound_appendix} has the form
\begin{equation}
  f(x) = \phi(x) - \phi \bigro{\sqrt{x^{2} + y^{2}}} \,,
\end{equation}
where we treat $y$ as a fixed parameter and $x$ should satisfy $x^{2} + y^{2}
\leq 1$. We show that this function is convex by lower bounding its second
derivative. First, the first and second derivatives of $\phi$ are
\begin{equation}
  \phi'(x) = - \frac{1}{2} \log_{2} \biggro{\frac{1 + x}{1 - x}}
\end{equation}
and
\begin{equation}
  \phi''(x) = - \frac{1}{\ln(2)} \frac{1}{1 - x^{2}} \,.
\end{equation}
Applying the product rule, the first and second derivatives of $f$ are
\begin{equation}
  f'(x) = \phi'(x) - \phi' \bigro{\sqrt{x^{2} + y^{2}}}
  \frac{x}{\sqrt{x^{2} + y^{2}}}
\end{equation}
and
\begin{IEEEeqnarray}{rCl}
  f''(x) &=& \phi''(x) - \phi'' \bigro{\sqrt{x^{2} + y^{2}}}
  \frac{x^{2}}{x^{2} + y^{2}} \IEEEnonumber \\
  &&-\> \phi' \bigro{\sqrt{x^{2} + y^{2}}}
  \frac{y^{2}}{(x^{2} + y^{2})^{3/2}} \,.
\end{IEEEeqnarray}
Using that $\ln \bigro{\frac{1 + \abs{x}}{1 - \abs{x}}} \geq 2 \abs{x}$, the
last term can be replaced with
\begin{equation}
  - \phi' \bigro{\sqrt{x^{2} + y^{2}}} \frac{y^{2}}{(x^{2} + y^{2})^{3/2}}
  \geq \frac{1}{\ln(2)} \frac{y^{2}}{x^{2} + y^{2}} \,,
\end{equation}
so that
\begin{IEEEeqnarray}{rCl}
  f''(x) &\geq& \frac{1}{\ln(2)}
  \begin{IEEEeqnarraybox*}[][t]{rl}
    \biggsq{&- \frac{1}{1 - x^{2}} \\
      &+\> \frac{1}{1 - x^{2} - y^{2}} \frac{x^{2}}{x^{2} + y^{2}}
      + \frac{y^{2}}{x^{2} + y^{2}}}
  \end{IEEEeqnarraybox*} \IEEEnonumber \\
  &=& \frac{1}{\ln(2)} \biggsq{- \frac{1}{1 - x^{2}}
      + \frac{1 - y^{2}}{1 - x^{2} - y^{2}}} \IEEEnonumber \\
  &=& \frac{1}{\ln(2)} \frac{x^{2} y^{2}}{(1 - x^{2})(1 - x^{2} - y^{2})}
  \IEEEnonumber \\
  &\geq& 0 \,,
\end{IEEEeqnarray}
which shows that $f$ is convex. Noticing that $f'(0) = 0$ (or just that $f$
is an even function) implies that $x = 0$ is the global minimum.

\subsection*{Proof of Lemma~\ref{lem:fidel_trnormW}}

A basic property of the trace norm is that $\trnorm{W_{\rB}} = \Tr[U_{\rB}
W_{\rB}]$ for some unitary operator $U_{\rB}$; furthermore, since $W_{\rB}$
is Hermitian, $U_{\rB}$ can also be taken to be Hermitian. From here and
using that $W = \trans{\alpha}{\alpha'} + \trans{\alpha'}{\alpha}$,
\begin{IEEEeqnarray}{rCl}
  \trnorm{W_{\rB}} &=& \Tr[U_{\rB} W_{\rB}] \IEEEnonumber \\
  &=& \Tr \bigsq{(U_{\rB} \otimes \id_{\rE}) W} \IEEEnonumber \\
  &=& 2 \re \bigsq{\bra{\alpha} U_{\rB} \otimes \id_{\rE} \ket{\alpha'}}
  \IEEEnonumber \\
  &\leq& 2 \babs{\bra{\alpha} U_{\rB} \otimes \id_{\rE} \ket{\alpha'}}
  \IEEEnonumber \\
  &\leq& 2 F(\alpha\0_{\rE}, \alpha\1_{\rE}) \,.
\end{IEEEeqnarray}
The final line follows, by Uhlmann's theorem, from noticing that
$\ket{\alpha}$ and $U_{\rB} \otimes \id_{\rE} \ket{\alpha'}$ are
purifications of $\alpha\0_{\rE}$ and $\alpha\1_{\rE}$.

\balance

\subsection*{Proof of Lemma~\ref{lem:fidel_trcoeffs}}

We introduce purifications $\ket{\chi_{0}}$ and $\ket{\chi_{1}}$ of
$\tau_{0}$ and $\tau_{1}$ such that $F(\tau_{0}, \tau_{1}) =
\braket{\chi_{0}}{\chi_{1}}$. In terms of these, note that
\begin{IEEEeqnarray}{rCl}
  \ket{\psi} &=& \sqrt{p_{0}} \ket{\chi_{0}} \ket{\gamma_{0}}
  + \sqrt{p_{1}} \ket{\chi_{1}} \ket{\gamma_{1}} \,,\\
  \ket{\phi} &=& \sqrt{q_{0}} \ket{\chi_{0}} \ket{\delta_{0}}
  + \sqrt{q_{1}} \ket{\chi_{1}} \ket{\delta_{1}} \,,
\end{IEEEeqnarray}
where $\{\ket{\gamma_{0}}, \ket{\gamma_{1}}\}$ and
$\{\ket{\delta_{0}}, \ket{\delta_{1}}\}$ are orthonormal bases, are
purifications of $\rho$ and $\sigma$. Using Uhlmann's theorem and expanding,
the fidelity between $\rho$ and $\sigma$ is lower bounded by
\begin{IEEEeqnarray}{rCl}
  F(\rho, \sigma) &\geq& \babs{\braket{\psi}{\phi}} \IEEEnonumber \\
  &=& \Babs{\sum_{ij} \sqrt{p_{i} q_{j}} \braket{\chi_{i}}{\chi_{j}}
    \braket{\gamma_{i}}{\delta_{j}}} \IEEEnonumber \\
  &=& \Babs{\sum_{ij} \sqrt{p_{i} q_{j}} F(\tau_{i}, \tau_{j}) U_{ji}}
  \IEEEnonumber \\
  &=& \babs{\Tr[U T]} \,,
\end{IEEEeqnarray}
where $U$ and $T$ are the matrices of elements
$U_{ji} = \braket{\gamma_{i}}{\delta_{j}}$ and
$T_{ij} = \sqrt{p_{i} q_{j}} F(\tau_{i}, \tau_{j})$. By exploiting the
freedom to choose the bases $\{\ket{\gamma_{0}}, \ket{\gamma_{1}}\}$ and
$\{\ket{\delta_{0}}, \ket{\delta_{1}}\}$, $U$ can be made to be any
$2 \times 2$ unitary matrix. Maximising the right-hand side over $U$, we
obtain
\begin{equation}
  F(\rho, \sigma) \geq \trnorm{T} \,,
\end{equation}
with
\begin{equation}
  \label{eq:T_fidels_def}
  T = \begin{bmatrix}
    \sqrt{p_{0} q_{0}} \trnorm{\tau_{0}}
    & \sqrt{p_{0} q_{1}} F(\tau_{0}, \tau_{1}) \\
    \sqrt{p_{1} q_{0}} F(\tau_{0}, \tau_{1})
    & \sqrt{p_{1} q_{1}} \trnorm{\tau_{1}}
  \end{bmatrix} \,,
\end{equation}
in which we inserted that $F(\tau_{i}, \tau_{i}) = \trnorm{\tau_{i}}$.

In general, the trace norm of a $2 \times 2$ matrix
$M = \bigsq{\begin{smallmatrix} \alpha & \beta \\ \gamma &
    \delta \end{smallmatrix}}$ is given by
\begin{equation}
  \trnorm{M} = \sqrt{T + 2 \sqrt{D}} \,,
\end{equation}
where
\begin{IEEEeqnarray}{rCl}
  T &=& \abs{\alpha}^{2} + \abs{\beta}^{2}
  + \abs{\gamma}^{2} + \abs{\delta}^{2} \,, \\
  \sqrt{D} &=& \babs{\alpha \delta - \beta \gamma}
\end{IEEEeqnarray}
are respectively the trace of $\abs{M}^{2} = M^{\dagger} M$ and the root of
its determinant. Applying this to obtain an explicit expression for the trace
norm of \eqref{eq:T_fidels_def} and using that
$F(\tau_{0}, \tau_{1}) \leq \sqrt{\trnorm{\tau_{0}}}
\sqrt{\trnorm{\tau_{1}}}$ produces the result
\begin{IEEEeqnarray}{rCl}
  F(\rho, \sigma)^{2} &\geq& \bigro{\sqrt{p_{0} q_{0}} \trnorm{\tau_{0}}
    + \sqrt{p_{1} q_{1}} \trnorm{\tau_{1}}}^{2} \IEEEnonumber \\
  &&+\> \bigro{\sqrt{p_{0} q_{1}} - \sqrt{p_{1} q_{0}}}^{2}
  F(\tau_{0}, \tau_{1})^{2} \,.
\end{IEEEeqnarray}

\end{document}